\documentstyle{mn}

\newif\ifAMStwofonts
\ifoldfss
  \ifCUPmtlplainloaded \else
    \NewTextAlphabet{textbfit} {cmbxti10} {}
    \NewTextAlphabet{textbfss} {cmssbx10} {}
    \NewMathAlphabet{mathbfit} {cmbxti10} {} % for math mode
    \NewMathAlphabet{mathbfss} {cmssbx10} {} %  "   "    "
  \fi
  \ifAMStwofonts
    \ifCUPmtlplainloaded \else
      \NewSymbolFont{upmath} {eurm10}
      \NewSymbolFont{AMSa} {msam10}
      \NewMathSymbol{\upi}     {0}{upmath}{19}
      \NewMathSymbol{\umu}     {0}{upmath}{16}
      \NewMathSymbol{\upartial}{0}{upmath}{40}
      \NewMathSymbol{\leqslant}{3}{AMSa}{36}
      \NewMathSymbol{\geqslant}{3}{AMSa}{3E}

    \fi
  \fi
\fi % End of OFSS

\ifnfssone
  \newmathalphabet{\mathit}
  \addtoversion{normal}{\mathit}{cmr}{m}{it}
  \addtoversion{bold}{\mathit}{cmr}{bx}{it}
  \newmathalphabet{\mathbfit} % math mode version of \textbfit{..}
  \addtoversion{normal}{\mathbfit}{cmr}{bx}{it}
  \addtoversion{bold}{\mathbfit}{cmr}{bx}{it}
  \newmathalphabet{\mathbfss} % math mode version of \textbfss{..}
  \addtoversion{normal}{\mathbfss}{cmss}{bx}{n}
  \addtoversion{bold}{\mathbfss}{cmss}{bx}{n}
  \ifAMStwofonts
    \ifCUPmtlplainloaded \else
      \UseAMStwoboldmath
      \makeatletter
      \new@mathgroup\upmath@group
      \define@mathgroup\mv@normal\upmath@group{eur}{m}{n}
      \define@mathgroup\mv@bold\upmath@group{eur}{b}{n}
      \edef\UPM{\hexnumber\upmath@group}
      \new@mathgroup\amsa@group
      \define@mathgroup\mv@normal\amsa@group{msa}{m}{n}
      \define@mathgroup\mv@bold\amsa@group{msa}{m}{n}
      \edef\AMSa{\hexnumber\amsa@group}
      \makeatother
      \mathchardef\upi="0\UPM19
      \mathchardef\umu="0\UPM16
      \mathchardef\upartial="0\UPM40
      \mathchardef\leqslant="3\AMSa36
      \mathchardef\geqslant="3\AMSa3E
    \fi
  \fi
\fi % End of NFSS release 1

\ifnfsstwo
  \DeclareMathAlphabet{\mathbfit}{OT1}{cmr}{bx}{it}
  \SetMathAlphabet\mathbfit{bold}{OT1}{cmr}{bx}{it}
  \DeclareMathAlphabet{\mathbfss}{OT1}{cmss}{bx}{n}
  \SetMathAlphabet\mathbfss{bold}{OT1}{cmss}{bx}{n}
  \ifAMStwofonts
    \ifCUPmtlplainloaded \else
      \DeclareSymbolFont{UPM}{U}{eur}{m}{n}
      \SetSymbolFont{UPM}{bold}{U}{eur}{b}{n}
      \DeclareSymbolFont{AMSa}{U}{msa}{m}{n}
      \DeclareMathSymbol{\upi}{0}{UPM}{"19}
      \DeclareMathSymbol{\umu}{0}{UPM}{"16}
      \DeclareMathSymbol{\upartial}{0}{UPM}{"40}
      \DeclareMathSymbol{\leqslant}{3}{AMSa}{"36}
      \DeclareMathSymbol{\geqslant}{3}{AMSa}{"3E}
    \fi
  \fi
\fi % End of NFSS release 2

\ifCUPmtlplainloaded \else
  \ifAMStwofonts \else % If no AMS fonts
    \def\upi{\pi}
    \def\umu{\mu}
    \def\upartial{\partial}
  \fi
\fi

\input{epsfig.sty}

\title{Higher order variability properties of accreting black holes}
\author[T. J. Maccarone \& P. S. Coppi]
       {Thomas J. Maccarone \\
        Scuola Internationale Superiore di Studi Avanzati, via Beirut, n. 2-4, Trieste, Italy, 34014 
	\newauthor
	Paolo S. Coppi\\
	Department of Astronomy, Yale University, P.O. Box 208101, New Haven CT USA 06520-8101}
	
\date{}

\pagerange{\pageref{firstpage}--\pageref{lastpage}}
\pubyear{}

\begin{document}

\maketitle

\label{firstpage}

\begin{abstract}
To better constrain the emission mechanism underlying the hard state
of galactic black hole candidates, we use high-time resolution RXTE
lightcurves for Cyg X-1 and GX 339-4 to compute two higher order
variability statistics for these objects, the skewness and the Fourier
bispectrum. Similar analyses, in particular using the skewness
measure, have been attempted before, but the photon collection area of
RXTE allows us to present results of much greater statistical
significance.  The results for the two objects are qualitatively
similar, reinforcing the idea that the same basic mechanisms are at
work in both. We find a significantly positive skewness for
variability timescales less than $\sim 1$ second, and a {\it negative}
skewness for timescales from $1- 5$ sec.  Such a skewness pattern
cannot be reproduced by the simplest shot variability models where
individual shots have a fixed profile and intensity and are
uncorrelated in time.  Further evidence against simple shot models
comes from the significant detection of a non-zero bicoherence for
Fourier periods $\sim 0.1-10$ sec, implying that significant coupling
does exist between variations on these timescales.  We discuss how
current popular models for variability in black hole systems can be
modified to match these observations.  Using simulated light curves,
we suggest that the most likely way to reproduce this observed
behavior is to have the variability come in groups of many shots, with
the number of shots per unit time fitting an envelope function which
has a rapid rise and slow decay, while the individual shots have a
slow rise and a rapid decay. Invoking a finite energy reservoir that
is depleted by each shot is a natural way of producing the required
shot correlations.

\end{abstract}


\section{Introduction} 
The continuum X-ray and $\gamma$-ray spectra of accreting black holes
generally consist of a soft component that can be described as a
multi-colored black body and a hard component that can be well fit by
a cutoff power law with some additional flux due to Compton reflection
(see e.g. Zdziarski et al.  1997; Gierlinski et al. 1997; Poutanen
1998).  Thermal Comptonisation models (with an additional non-thermal
component sometimes required) provide a physical basis for producing
such a spectrum.  A variety of models with differing geometries can
fit the spectral data (see e.g.  Poutanen 1998; Beloborodov 1999;
Zdziarski 2000), so one must find a means of breaking the spectral
degeneracies in order to differentiate between possible geometries of
the flow and mechanisms for producing the high energy electrons that
upscatter photons into the hard tail of the spectrum.

Another useful characteristic of accreting black holes, especially
those in hard spectral states, is rapid variability (see e.g. Nolan et
al.  1981; Negoro, Miyamoto \& Kitamoto 1994; Vaughan \& Nowak 1997).
Several variability properties of their lightcurve have already been
studied extensively, especially for Cygnus X-1, and place constraints
on the underlying emission mechanism.  For example, Fourier spectra
and cross-spectra have been used to show that the spectral energy
distributions of Cygnus X-1 and GX 339-4 (the two canonical persistent
black hole candidates in the Milky Way) cannot be produced by Compton
scattering in a static, uniformly dense cloud of hot electrons
(Miyamoto et al.  1988), as had been previously suggested (Payne
1980).  The improved photon statistics that are now possible with
large area detectors like the Rossi X-ray Timing Explorer (RXTE) merit
a re-examination of some of these prior analyses.

In a previous paper, we presented measurements of the
cross-correlation function and autocorrelation functions of Cygnus X-1
at different energies which rule out all models in which the time lags
between energy bands come from diffusion timescales through the
corona.  We also found that the autocorrelation and cross-correlation
functions imply that the shape of individual shots in the hard state
of Cygnus X-1 is such that the count rates rise more or less
exponentially, then decay much more rapidly (Maccarone, Coppi \&
Poutanen 2000), the inverse of the FRED (fast-rise, exponential decay)
behavior often cited in $\gamma$-ray burst studies (see e.g. Fenimore,
Madras \& Nayakshin 1996).  This possibility can be evaluated and
understood further by using the time skewness statistic, which
measures the symmetry of a lightcurve as a function of time scale.
Previous time skewness analyses showed that the light curve of Cygnus
X-1 cannot be produced by purely exponential shots with either a
rising or falling profile (Priedhorsky et al. 1979), but in our
knowledge, this technique has not been applied since, despite immense
advances in temporal resolution, photon statistics, and duration of
observations. Here we present the results of time skewness analyses of
RXTE data sets from Cygnus X-1 and GX 339-4 in their hard states.  To
complement this analysis, we also compute the Fourier bispectrum of
the lightcurve data.  The Fourier bispectrum (e.g. Mendel 1991) has
been applied the spatial properties of the cosmic microwave background
and large-scale structure (e.g. Cooray 2001) and to problems in other
fields such as neurobiology (e.g. Bullock et al. 1995) and speech
recognition (Fackrell 1996), but it has not previously been applied to
astronomical time series data.  With good quality data, however, we
find that it provides another useful, higher order variability probe,
determining whether the fluctuations on one time scale are coupled to
the fluctuations on other timescales.

In section 2 we briefly describe the data set used and the data
reduction and screening.  In section 3, we define the statistics we
use and present the observational results.  In section 4, we discuss
the mathematical form of the light curve implied by the observational
results and consider the possible physical models that could produce
the observed variability.

\section{OBSERVATIONS} 

\subsection{Reduction}

Because of the need for large photon numbers to compute higher order
variability statistics, we analyze the RXTE hard state data for the
two brightest quasi-persistent low/hard state sources, Cyg X-1 and GX
339-4. The observations used are listed in Table 1.  Light curves are
extracted from the Proportional Counter Array (PCA) data in several
different energy bands using the standard RXTE screening criteria of
earth elevation greater than 10 degrees, pointing offset less than
0.01 degrees, having all 5 proportional counter units on, and having
the standard amounts of time before and after South Atlantic Anomaly
passages. In all cases, the time skewness is computed on a 0.125
second time scale. The photon statistics do not permit a computation
on shorter timescales.

\begin{table*}
\centering
\caption{The data analyzed for this the time skewness analyses.  Dates
are presented in DD/MM/YY format. }
\begin{tabular}{@{}lccc@{}}
ObsID & Start Time & Stop Time & Source\\ 
30158-01-01-00& 10/12/97 07:07:53& 10/12/97 08:30:14 & Cyg X-1\\
30158-01-02-00& 11/12/97 07:06:14& 11/12/97 08:45:14 & Cyg X-1\\
30158-01-03-00& 14/12/97 08:48:14& 14/12/97 10:20:14 & Cyg X-1\\
30158-01-04-00& 15/12/97 03:49:42& 15/12/97 05:26:14 & Cyg X-1\\
30158-01-05-00& 15/12/97 05:26:14& 15/12/97 07:09:14 & Cyg X-1\\
30158-01-06-00& 17/12/97 00:39:55& 17/12/97 02:05:14 & Cyg X-1\\
30158-01-07-00& 20/12/97 07:11:18& 20/12/97 08:29:14 & Cyg X-1\\
30158-01-08-00& 21/12/97 05:28:14& 21/12/97 07:05:14 & Cyg X-1\\
30158-01-09-00& 24/12/97 23:03:14& 24/12/97 00:39:14 & Cyg X-1\\
30158-01-10-00& 25/12/97 00:39:14& 25/12/97 01:45:14 & Cyg X-1\\
30158-01-11-00& 30/12/97 02:18:03& 30/12/97 03:52:14 & Cyg X-1\\
30158-01-12-00& 30/12/97 03:52:14& 30/12/97 05:30:14 & Cyg X-1\\
20181-01-01-01& 03/02/97 15:56:24& 03/02/97 19:09:13 & GX 339-4 \\
20181-01-01-00& 03/02/97 22:27:01& 04/02/97 01:36:13 & GX 339-4 \\
20181-01-02-00& 10/02/97 15:49:19& 10/02/97 20:51:13 & GX 339-4 \\
20181-01-03-00& 17/02/97 18:28:38& 18/02/97 00:12:13 & GX 339-4 \\
20183-01-01-00& 08/02/97 14:20:30& 08/02/97 20:48:13 & GX 339-4 \\
20183-01-02-00& 14/02/97 00:09:12& 14/02/97 06:43:13 & GX 339-4 \\
20183-01-02-01& 14/02/97 14:20:45& 14/02/97 21:22:13 & GX 339-4 \\
20183-01-03-00& 22/10/97 03:00:35& 22/10/97 05:52:14 & GX 339-4 \\
20183-01-04-00& 25/10/97 03:22:40& 25/10/97 06:16:14 & GX 339-4 \\
20183-01-05-00& 28/10/97 18:07:56& 28/10/97 22:13:14 & GX 339-4 \\
20183-01-06-00& 31/10/97 19:40:12& 31/10/97 22:10:14 & GX 339-4 \\
20183-01-07-00& 03/11/97 20:35:02& 03/11/97 23:48:14 & GX 339-4 \\
\end{tabular}
\end{table*}

\begin{table*}
\begin{center}
\caption{The data analyzed for the bicoherence analyses.  Dates are
presented in DD/MM/YY format. }
\begin{tabular}{lccc}
ObsID & Start Time & Stop Time & Source\\ 30158-01-01-00& 10/12/97
07:07:53& 10/12/97 08:30:14 & Cyg X-1\\ 30158-01-02-00& 11/12/97
07:06:14& 11/12/97 08:45:14 & Cyg X-1\\ 30158-01-03-00& 14/12/97
08:48:14& 14/12/97 10:20:14 & Cyg X-1\\ 30158-01-04-00& 15/12/97
03:49:42& 15/12/97 05:26:14 & Cyg X-1\\ 30158-01-05-00& 15/12/97
05:26:14& 15/12/97 07:09:14 & Cyg X-1\\ 30158-01-06-00& 17/12/97
00:39:55& 17/12/97 02:05:14 & Cyg X-1\\ 30158-01-07-00& 20/12/97
07:11:18& 20/12/97 08:29:14 & Cyg X-1\\ 30158-01-08-00& 21/12/97
05:28:14& 21/12/97 07:05:14 & Cyg X-1\\ 30158-01-09-00& 24/12/97
23:03:14& 24/12/97 00:39:14 & Cyg X-1\\ 30158-01-10-00& 25/12/97
00:39:14& 25/12/97 01:45:14 & Cyg X-1\\ 30158-01-11-00& 30/12/97
02:18:03& 30/12/97 03:52:14 & Cyg X-1\\ 30158-01-12-00& 30/12/97
03:52:14& 30/12/97 05:30:14 & Cyg X-1\\ 20056-01-01-00& 05/04/97
08:36:14& 05/04/97 09:15:13 & GX 339-4\\ 20056-01-02-00& 10/04/97
11:47:29& 10/04/97 12:28:13 & GX 339-4\\ 20056-01-03-00& 11/04/97
13:25:44& 11/04/97 14:06:13 & GX 339-4\\ 20056-01-04-00& 13/04/97
20:09:51& 13/04/97 20:50:13 & GX 339-4\\ 20056-01-05-00& 15/04/97
20:41:10& 15/04/97 21:23:13 & GX 339-4\\ 20056-01-06-00& 17/04/97
23:25:47& 18/04/97 00:01:13 & GX 339-4\\ 20056-01-06-01& 18/04/97
00:01:13& 18/04/97 00:09:13 & GX 339-4\\ 20056-01-07-00& 19/04/97
22:20:01& 19/04/97 23:21:13 & GX 339-4\\ 20056-01-08-00& 22/04/97
21:53:21& 22/04/97 22:29:13 & GX 339-4\\

\end{tabular}
\end{center}
\end{table*}

\section{ANALYSIS}
\subsection{Time Skewness}
\subsubsection{Definition}
The time skewness measures the asymmetry of a lightcurve.  We choose a
slightly different definition than Priedhorsky et al. (1979) in order
to non-dimensionalise the skewness and to make the normalisation
independent of constant scaling factor changes to the count rate:
\begin{equation}
TS(\tau) = \frac{1}{\sigma^3} \overline{ [(s_i-\bar{s})^2(s_{i-u}-\bar{s})-(s_i-\bar{s})(s_{i-u}-\bar{s})^2]},
\end{equation}
where $\tau$ is defined to be $u$ times the time bin size, $s_i$ is
count rate for the $i$th element of the light curve, and $\sigma$ is
the standard deviation of the count rate.  Essentially, the time
skewness is a weighted average of the difference between two count
rates measured at a given separation in time.  If a light curve is
composed of the summation of spikes, and the spikes typically rise
more sharply than they fall on a particular time scale, the sign of
the skewness will be negative; if the spikes fall more sharply than
they rise, the skewness will be positive.  We compute the time
skewness for each observation separately, then present the weighted
average of the individual measurements.  The errors plotted are the
rms variations of the individual measurements used in the averaging,
and may be overestimated if there is intrinsic variation in the time
skewness.

\subsubsection{Results}

Throughout the rest of this paper the term ``shot'' will be used to
describe a temporary increase in luminosity.  In some cases in the
literature this term has been used to describe models where the
lightcurve consists of the random superposition of events with a
single profile, duration, and intensity.  Except when explicitly
stated to have this meaning, we intend a more general definition of
the word, and by referring to ``shots'', we may imply a distribution
of flare shapes and durations.  Additionally, the term ``rising shot''
will be used to describe a shot whose duration is dominated by its
rise time, and the term ``falling shot'' will be used to describe a
shot whose duration is dominated by its decay time.

The 2-5 keV (channels 0-13 in the PCA) skewnesses from Cygnus X-1 and
GX 339-4 in their hard states are presented in Figure 1.  Both objects
have a strongly positive skewness out to about 1 second and a strongly
negative skewness between 1 second and the time scale on which their
autocorrelation functions go to zero (5 seconds for Cygnus X-1 and 3
seconds for GX339-4).  Beyond the autocorrelation function decay time
scale, their skewnesses are consistent with zero.  Whether these
characteristics apply to all objects' hard states is difficult to
determine, because the other persistent hard state sources are much
fainter than Cyg X-1 and GX 339-4.  Furthermore, the transient hard
state sources often show strong, relatively narrow QPOs in their hard
states which complicate this analysis.

The skewnesses for other energy bands were also computed and are
presented in Figure 2, with the error bars removed to make the figure
more clear.  In Cygnus X-1, the skewnesses are identical within the
errors across energy bands.  There seems to be some indication for a
more symmetric overall lightcurve (i.e. a lower amplitude of the
skewness) on long timescales in the higher energy bands, but it is not
clear whether this is a real effect or due to the lower count rates at
higher energy introducing more Poisson noise. In GX 339-4, the
skewnesses become more symmetric as a function of increasing energy.
This agrees well with the observation that the Fourier spectrum in GX
339-4 is consistent with being constant as a function of energy, while
the Fourier time lags become longer for larger energy separations
(e.g. Lin et al. 2000).  We note that the large difference between the
15-45 keV lightcurve's skewness and the lower energy bands' skewnesses
is due to the effects of Poisson noise artifically increasing the
variance which is used to normalise the skewness values.  This cannot
be the case for the two lower energy bands, since the intermediate
band has a count rate about 10\% {\it higher} than the lowest energy
band, and still has a lower skewness.  Still the differences in the
skewness as a function of energy are not particularly large, and it
would be worthwhile to re-investigate this effect with the entire RXTE
data set for GX 339-4 at the end of the mission.

The combination of the skewness measurements and the previous power
spectrum results suggests that as the energy band increases, the rise
timescale of a shot becomes slightly shorter while its decay timescale
becomes somewhat longer.  Thus the shots in GX 339-4 become more
symmetric at higher energies and the normalisation of the skewness
becomes smaller.  Previous results found only that one-sided
exponential shot models could be ruled out, and did not look to long
enough timescales to see the sign change in the skewness (Priedhorsky
et al. 1979).  The fluctuations in the amplitude of the negative peak
are of order the fluctutations in the skewness on 5-10 second
timescales where the mean observed and expected values are $\sim 0$.
We thus do not ascribe any physical meaning to the observed
fluctations in the skewness within the peak.

The statistical properties of the time skewness are not well
established.  In order to ensure that our results are not due to
meaaurement errors, we create simulated data sets of time-symmetric
red noise with the phases at different frequencies uncorrelated with
one another, using the algorithm of Timmer \& K\"onig (1995), with a
power spectral form similar to the observed power spectrum of Cygnus
X-1 (a constant value below 0.1 Hz joined continuously to a $f^{-1}$
power law from 0.1 to 1 Hz and then to a $f^{-1.7}$ power law above 1
Hz, with an rms amplitude of 30\%).  One hundred lightcurves are
simulated, each with a duration of 10 kiloseconds (shorter than the 30
kilosecond integration used in the Cygnus X-1 data set, which means
that the real computation should be more robust than the simulated
data computation).  None of the simulated light curves show skewnesses
on time scales of less than 1 second that could match the first peak
(i.e. with absolute values greater than 0.02, the magnitude of the
first peak).  While three of the lightcurves show a skewness in the
2-4 second range with an absolute value of at least 0.04 (the size of
the negative peak), there are no lightcurves with absolute values
greater than 0.02 for the entire 2-4 second region, as is seen in
Cygnus X-1.  The mean value in the 2-4 second range is about 0.01,
which are about the same size as the errors plotted in Figure 1.
Thus, individual time bins from random light curves may occasionally
(but infrequently) show skewnesses as large as the observed one, but
they do not produce broad peaks in the skewness with depths as large
as the observed depths.  The randomly generated lightcurves do
frequently have skewnesses half the size of those observed.  This
underscores the importance of long integrations for making skewness
measurements, even when the photon statistics are good, since a short
lightcurve, even with perfect photon statistics, can produce large
skewness values despite having random Fourier phases at all
frequencies.

The power spectrum for GX 339-4 is roughly the same as for Cygnus X-1,
except that the high frequency cutoff occurs at about 3 Hz rather than
1 Hz.  We simulate 40 kilosecond lightcurves (our observations are 46
kiloseconds here) with the power spectral shape for GX 339-4.  As the
observed skewness for GX 339-4 has a broad trough from 3 to 5 seconds
at -0.015, we look for observations where the skewness has an absolute
value of at least 0.015 over the entire 3 to 5 second range.  We find
no such lightcurves out of our 100 simulated lightcurves, and find
only 3 lightcurves with any points with absolute values greater than
0.015 for any bin from 3 to 5 seconds.  Thus again we find that while
rarely, but occasionally, the depths of the troughs can be matched by
the red noise lightcurves, troughs with the necessary depth and
breadth cannot.

\begin{figure*}
\vskip 2 cm
\centerline{\epsfxsize=6.2cm \epsfysize=5.33cm \epsfbox{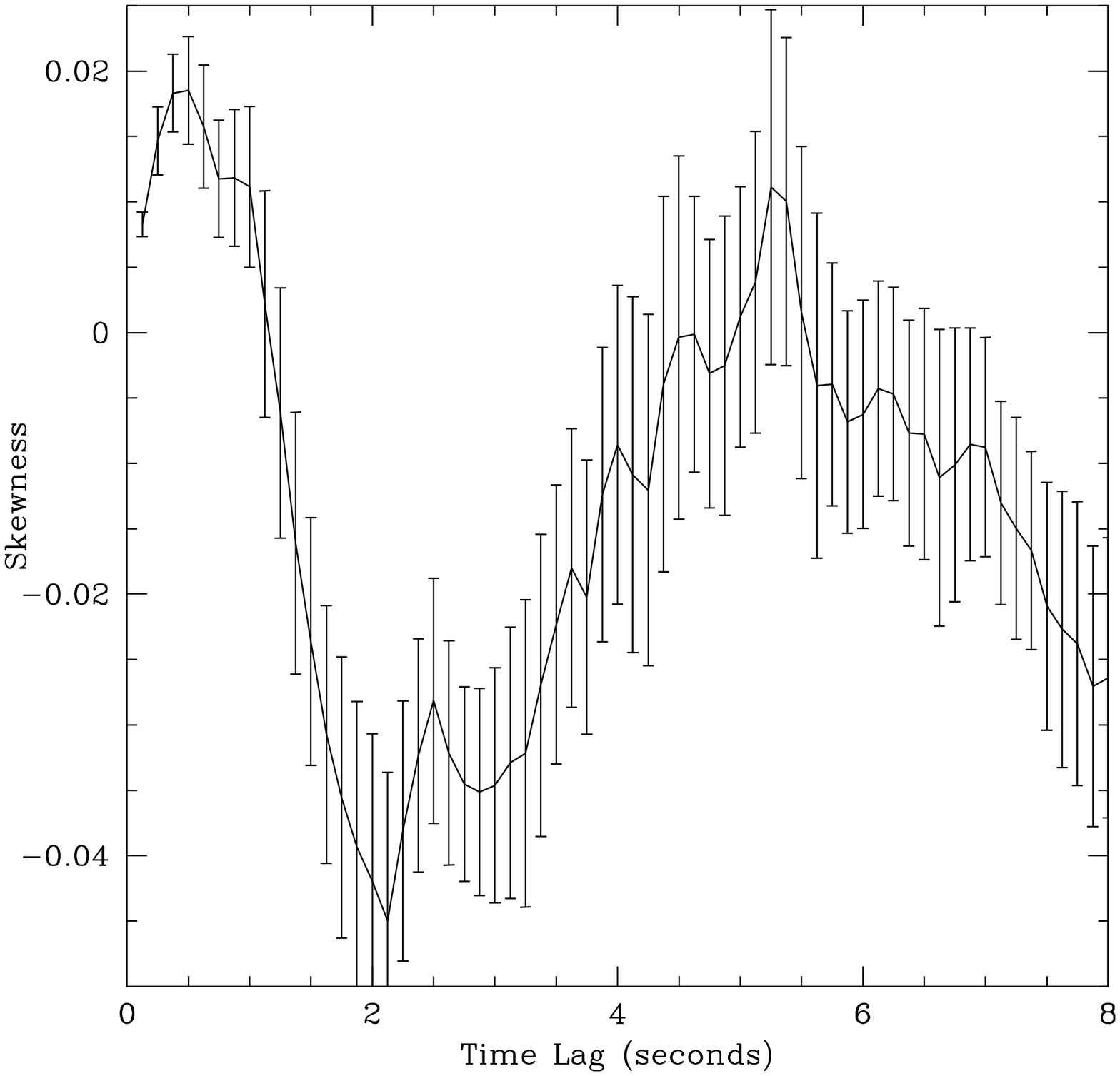}
\epsfxsize=6.2cm  \epsfysize=5.33cm \epsfbox{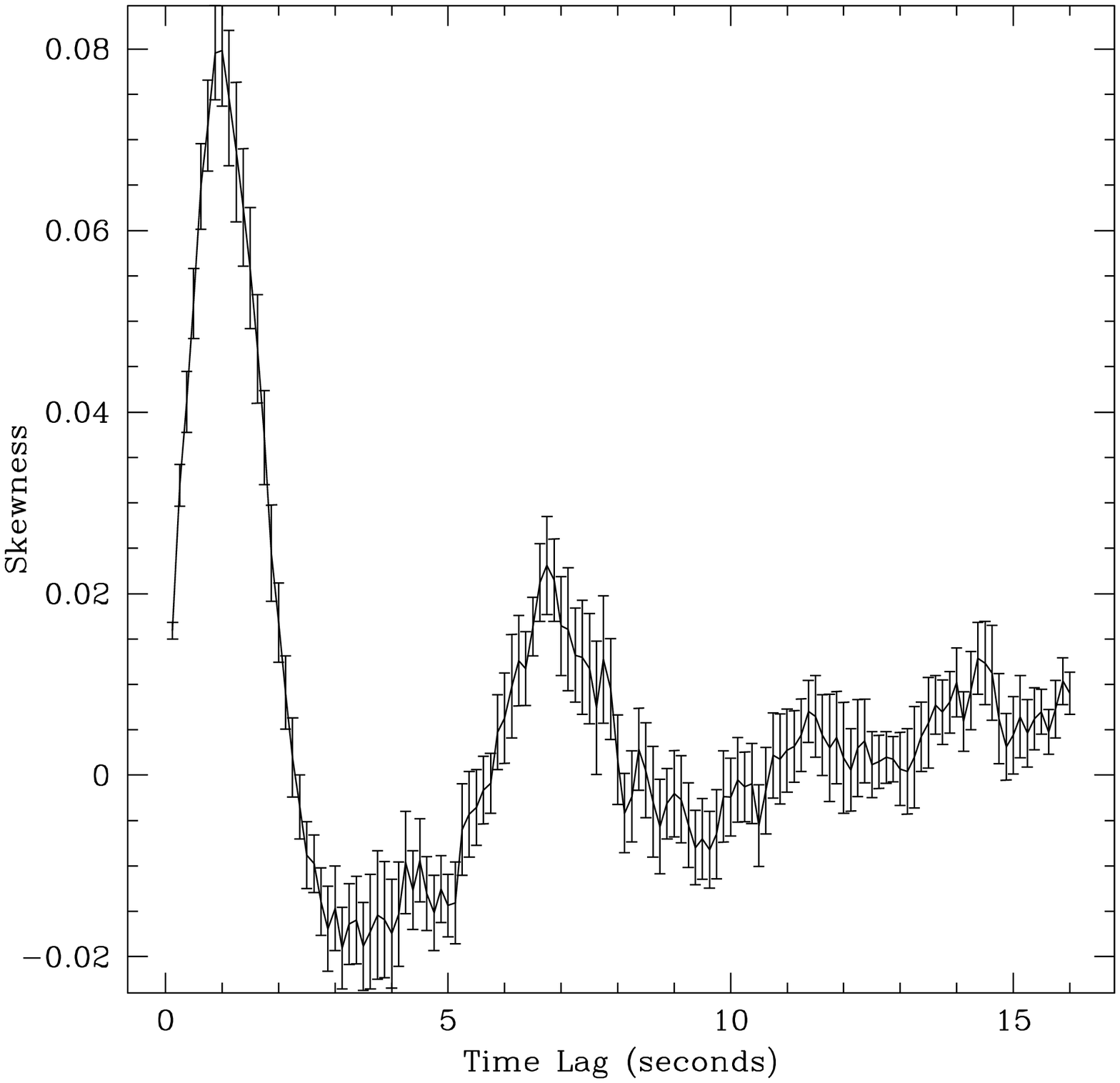}}
\caption {The 2-5 keV skewnesses for (a) the hard state of Cygnus X-1
and (b) the hard state of GX 339-4.}
\label{fig:hard}
\end{figure*}

\begin{figure*}
\vskip 2 cm
\centerline{\epsfxsize=6.2cm \epsfysize=5.33cm \epsfbox{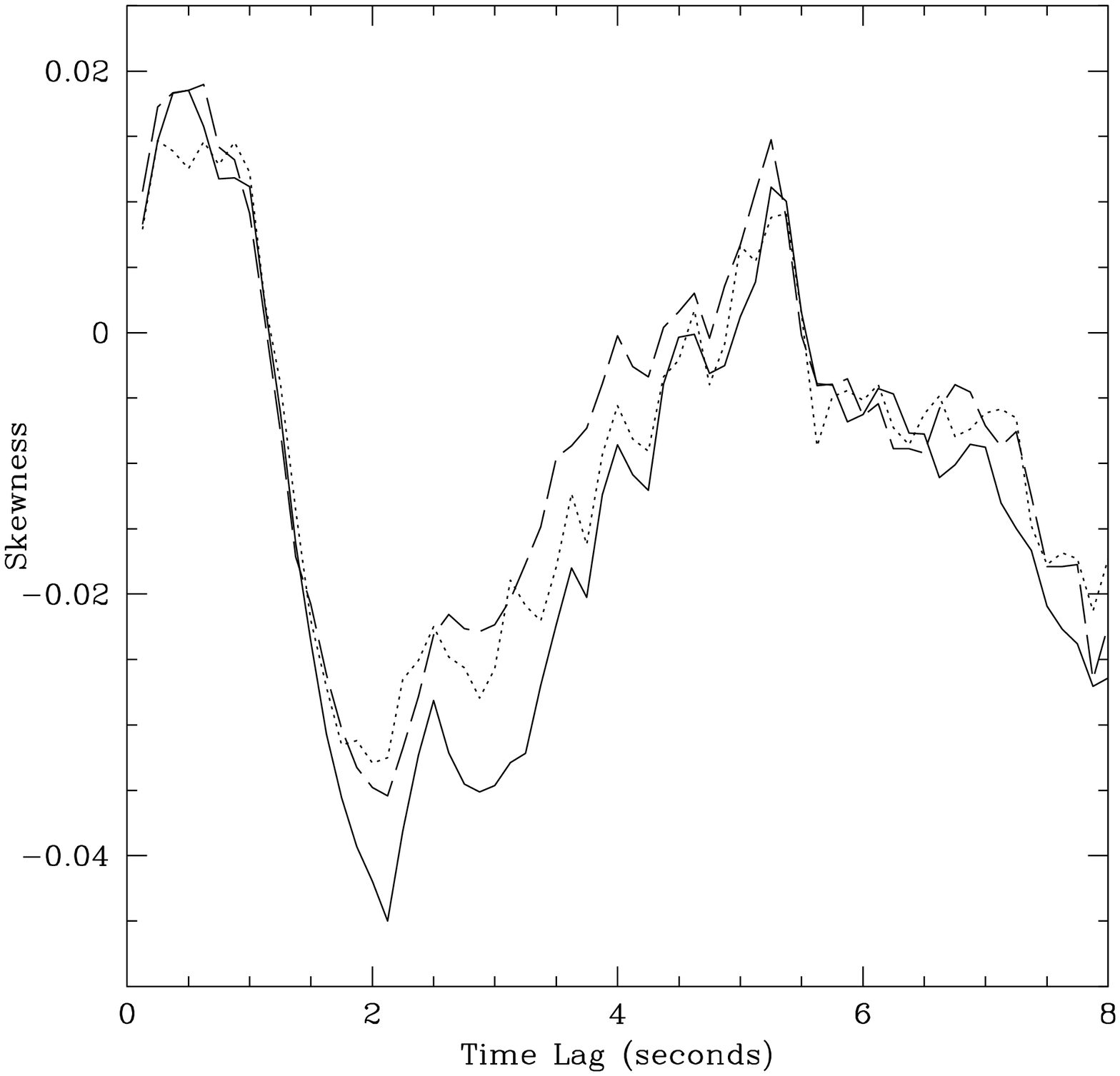}
\epsfxsize=6.2cm  \epsfysize=5.33cm \epsfbox{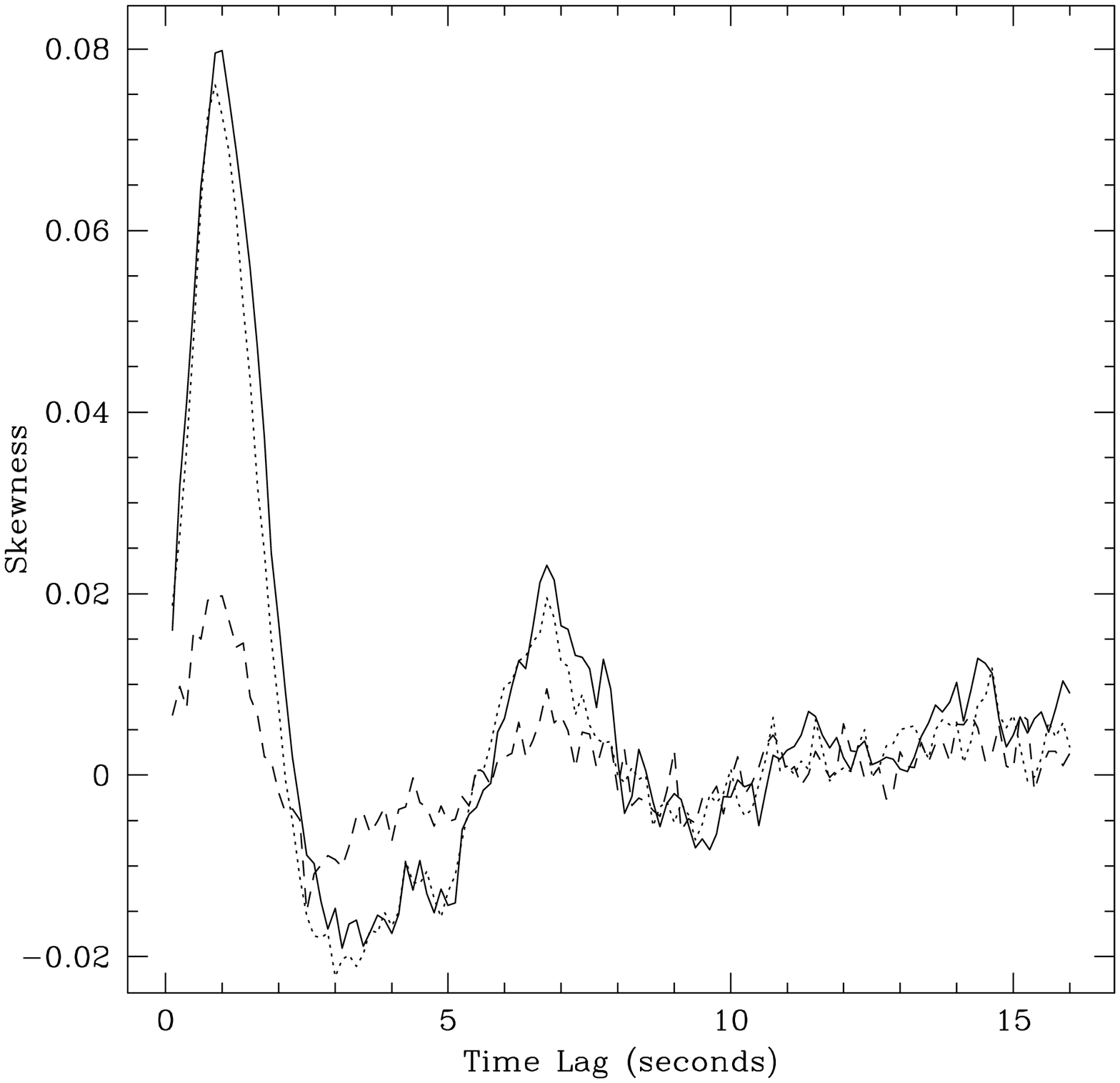}}
\caption {The higher energy band skewnesses, plotted without error
bars for (a) Cygnus X-1 in the hard state (solid line is 2-5 keV,
dotted line is 5-7 keV and dashed line is 7-10 keV) and (b) GX 339-4
in the hard state (solid line is 2-5 keV, dotted line is 5-15 keV, and
the dashed line is 15-45 keV).}
\end{figure*}

\subsubsection{Other spectral states}

The  time  skewnesses  were  also  computed  for  Cygnus  X-1  in  its
transition state and  in its soft state and for  LMC X-1 which appears
to have  only a soft state.   Since the rms  variability amplitudes in
the softer  states are  typically much lower  than in the  hard states
(making  variability  harder   to  measure),  and  the  characteristic
variability time  scales are  shorter (making binning  the data  up to
longer  time scales  problematic), the  skewnesses computed  were very
noisy, and it was difficult  to draw any useful conclusions from them.
Thus we neither plot nor discuss these results.

\subsection{Fourier bispectrum}

\subsubsection{Definition}
The bispectrum computed from a time series broken into $K$ segments is
defined as:
\begin{equation}
B(k,l)=\frac{1}{K} \sum_{i=0}^{K-1} X_i(k)X_i(l)X^*_i(k+l),
\end{equation}
where $X_i(k)$ is the frequency $k$ component of the discrete Fourier
transform of the $i$th time series (e.g. Mendel 1991; Fackrell 1996
and references within).  It is a complex quantity that measures the
magnitude and the phase of the correlation between the phases of a
signal at different Fourier frequencies.  Its value is unaffected by
additive Gaussian noise, although its variance will increase for a
noisy signal.

A related quantity, the bicoherence is the vector magnitude of the
bispectrum, normalised to lie between 0 and 1.  Defined analogously to
the coherence function (e.g. Nowak \& Vaughan 1996), it is the vector
sum of a series of bispectrum measurements divided by the sum of the
magnitudes of the individual measurements.  If the biphase (the phase
of the bispectrum) remains constant over time, then the bicoherence
will have a value of unity, while if the phase is random, then the
bicoherence will approach zero in the limit of an infinite number of
measurements.  The bicoherence is defined as:

\begin{equation}
b^2(k,l) = \frac{\left|\sum{X_i(k)X_i(l)X^*_i(k+l)}\right|^2}{\sum{\left|X_i(k)X_i(l)\right|^2}\sum{\left|X_i(k+l)\right|^2}}.
\end{equation}

This analysis will concentrate on the bicoherence, since statistically
significant measurements of it are easier to make.  The denominator of
this expression has a dependence on the amount of Gaussian noise, so a
correction will have to be made when the signal is noisy.
Additionally, the bicoherence, being bound between 0 and 1, has a
non-zero mean value due to errors even when the phases of the Fourier
transforms are uncorrelated.  The mean value and the standard
deviation of a bicoherence estimate based on $K$ distinct Fourier
transforms are both $\frac{1}{K}$.  The properties of the bispectrum
and the bicoherence as they relate to time series analysis are
well-reviewed by Fackrell (1996).

\subsubsection{Computation method and results}

The lightcurves are split into segments of 4096 elements (since the
time resolution of the lightcurves varies, the lengths of these
segments in absolute amounts of time will also vary).  Fourier
transforms are computed for each element.  The results are then
combined according to equation (2) to estimate the bicoherence.  The
data used to compute the bispectra are listed in Table 2.  

Since the bicoherence is a function of two independent variables, one
could presumably gain the most information from a 3 dimensional plot,
or a two dimensional contour plot.  Since we find from inspection
that, for these data, no individual peaks stand out in such plots, the
data are rebinned to present the results only as a function of the sum
of the two lower Fourier frequencies.  This allows for a simple
two-dimensional plot, allows for significant data rebinning to
increase the statistical significance of the results, and does not
remove any obvious underlying trends.  Similar techniques are
frequently used in analyses of spatial bispectra of the microwave
background, although in that case the computations are done in terms
of spherical harmonics rather than Fourier frequencies. We caution the
reader who may wish to try this technique on other time series that
for systems with strong, relatively narrow quasi-periodic
oscillations, rebinning the frequencies to one dimension may result in
the loss of important signals.  For example, we have examined a few
observations of GRS 1915+105 where the 1-10 Hz QPOs show harmonic
structure and have found that the bicoherence can be quite strong for
the pairs of frequencies that add up to the frequencies of the higher
harmonics, but that this effect can be seen clearly only from a three
dimensional plot.

We have plotted the results for the bicoherence measurements for
Cygnus X-1 (from 2-5 keV; channels 0 to 13) and GX 339-4 (from 2 to 8
keV; channels 0 to 25) in Figures 3 (a) and (b) respectively.  In all
cases, the bispectrum fits a broken power law with a negative index
for high frequencies; for lower frequencies, it has a nearly constant
value.  The non-zero bicoherence indicates that the variability at low
frequencies is coupled to the noise at high frequencies, providing an
additional piece of strong evidence that a single shot model cannot
match the data.  Furthermore the phases of the Fourier components are
more strongly correlated at low frequency than they are at high
frequency.

We attempted similar analyses for data from accreting neutron stars,
but the photon statistics were not sufficient to find interesting
results.  There is not sufficient RXTE data on any single neutron star
in the island state of an atoll source (the analog of the hard state
for black hole sources) to make use of this technique.  Likewise, LMC
X-1 (the most variable and best observed soft state source) lacks
sufficient signal to noise to make use of the bispectrum.  For objects
with strong quasi-periodic oscillations, the analysis is significantly
more complicated, as some frequencies may represent the harmonics of
periodicities found at lower frequencies.  Such analysis is outside
the scope of this paper which aims merely to characterise random noise
components of the power spectra of accreting objects.

\begin{figure*}
\vskip 3 cm
\centerline{\epsfxsize=9.3cm \epsfysize=8.0cm \epsfbox{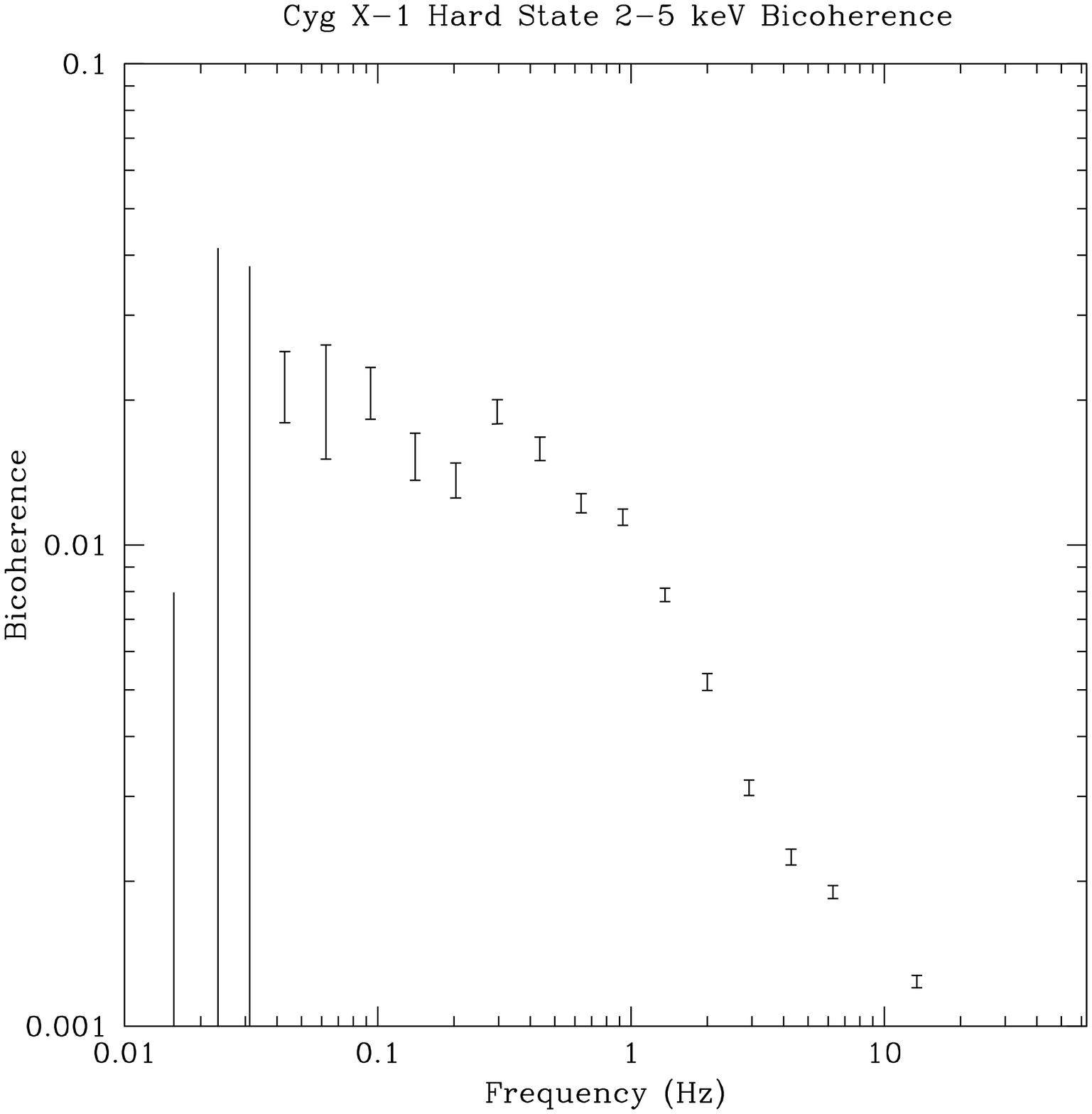}
\epsfxsize=9.3cm  \epsfysize=8.0cm \epsfbox{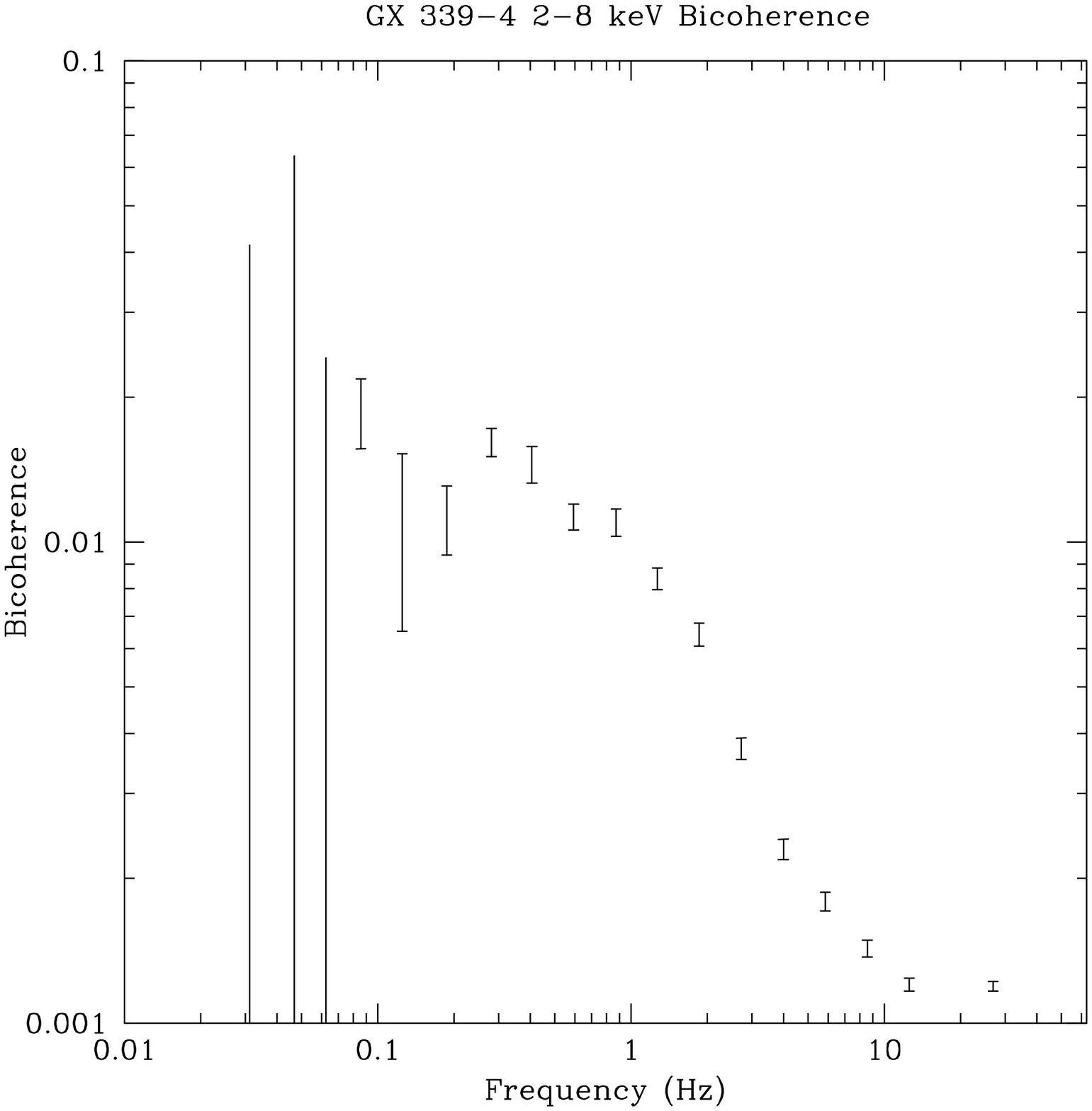}}
\caption{The bicoherence plots for (a) Cyg X-1 and (b) GX 339-4 in the
hard state.}
\end{figure*}

\section{Discussion} 

We now consider possible models and attempt to determine how they
match the key observational results - that (1) the skewness is
positive on short timescales, (2) the skewness changes sign and
becomes negative on timescales of about 1 to 5 seconds before going to
zero on long timescales and (3) that the bicoherence has a ``shelf''
at low bifrequency.  We also will make use of the result of
Priedhorsky et al. (1979), reproduced here, that the normalisation of
the skewness found in the data is much smaller than the normalisation
of the skewness found from a one-sided exponential shot distribution.
An exact reproduction of the time skewness is likely to be a quite
complicated problem numerically, as the shot shapes and arrival times
may take non-standard functional forms.  We first discuss toy models
that approximate the mathematical form of the skewness then discuss
recent theoretical models which might be expected to produce this
mathematical form.  Finally, we show that the bicoherence results
agree with the general mathematical form used to match the skewness,
providing an independent confirmation of our picture.  In the models
we consider, we invoke the ``shot noise'' hypothesis - that the
variability consists of discrete events.  Our conclusions that the
lightcurve shows a different sense of asymmetry on different
timescales are independent of this assumption.  The specific
interpretations below are not so clearly robust.  

One possible means of producing a skewness function which has both
positive and negative signs is to create a light curve which has both
rising and falling shots, with different time scales for the rising
shots than for the falling ones.  This would seem to imply two
physical mechanisms for producing shots.  Furthermore, to produce hard
lags on all time scales from this distribution, as is required from
the Fourier cross spectrum, one would need the rising shots to be
shorter with increasing energy while the falling shots would have to
be longer with increasing energy.  This would produce an
autocorrelation function that becomes wider with increasing energy on
long time scales, in violation of observational constraints.  We were
unable to find a distribution of shot intensities and time scales for
this scenario that reproduced the observed skewness.

A second model one might consider would be to have a random
superposition of identical two-sided shots, with a rise time different
from the fall time. Two sided shots, however, fail to reproduce to
observed skewness measurements.  They are always dominated by the
longer timescale, and actually fail to produce skewnesses which have
both substantial positive and negative portions on different time
scales.

A third possibility is to have all the shots be rising exponentials
with sharp cutoffs (inverse FREDs), while the probability of having a
shot occur at a given time is given by an exponentially falling
function (an actual FRED, i.e., the mean shot rate is proportional to
$\lambda \propto \exp[-(t-t_0)/\tau]$ for $(t>t_0)$. Thus the rising
shots are embedded in a falling envelope.  In this paper, the
envelopes will be modeled by drawing the arrival times of the peaks
of the shots from an exponential deviate; more complicated envelope
functions are possible but are not considered here.  Figure 4(a)
illustrates the difference between rising and falling shots, while
Figure 4(b) shows rising shots embedded in a falling envelope.

\begin{figure*}
\vskip 3 cm
\centerline{\epsfxsize=6.2cm \epsfysize=5.33cm \epsfbox{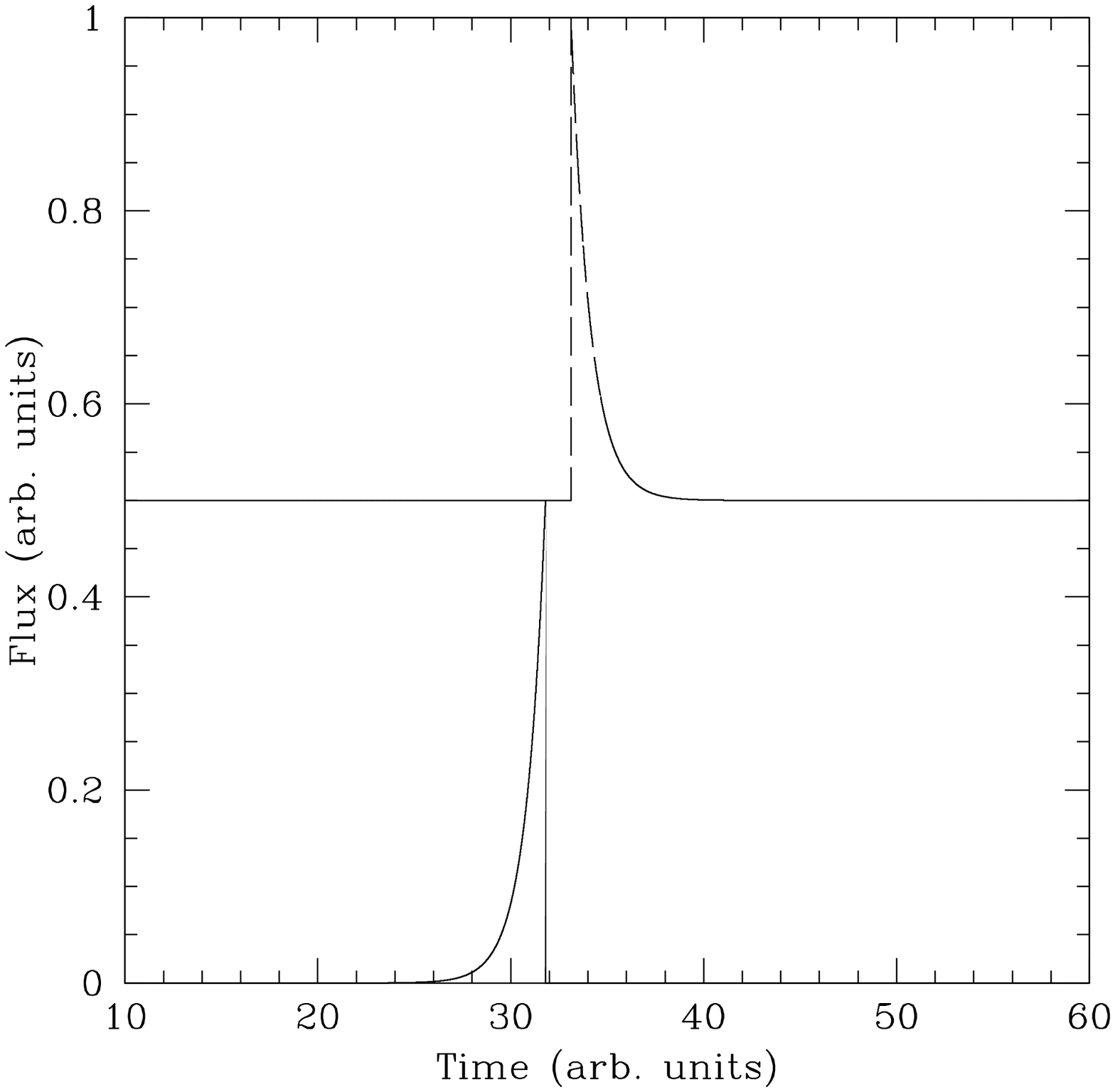}
\epsfxsize=6.2cm  \epsfysize=5.33cm \epsfbox{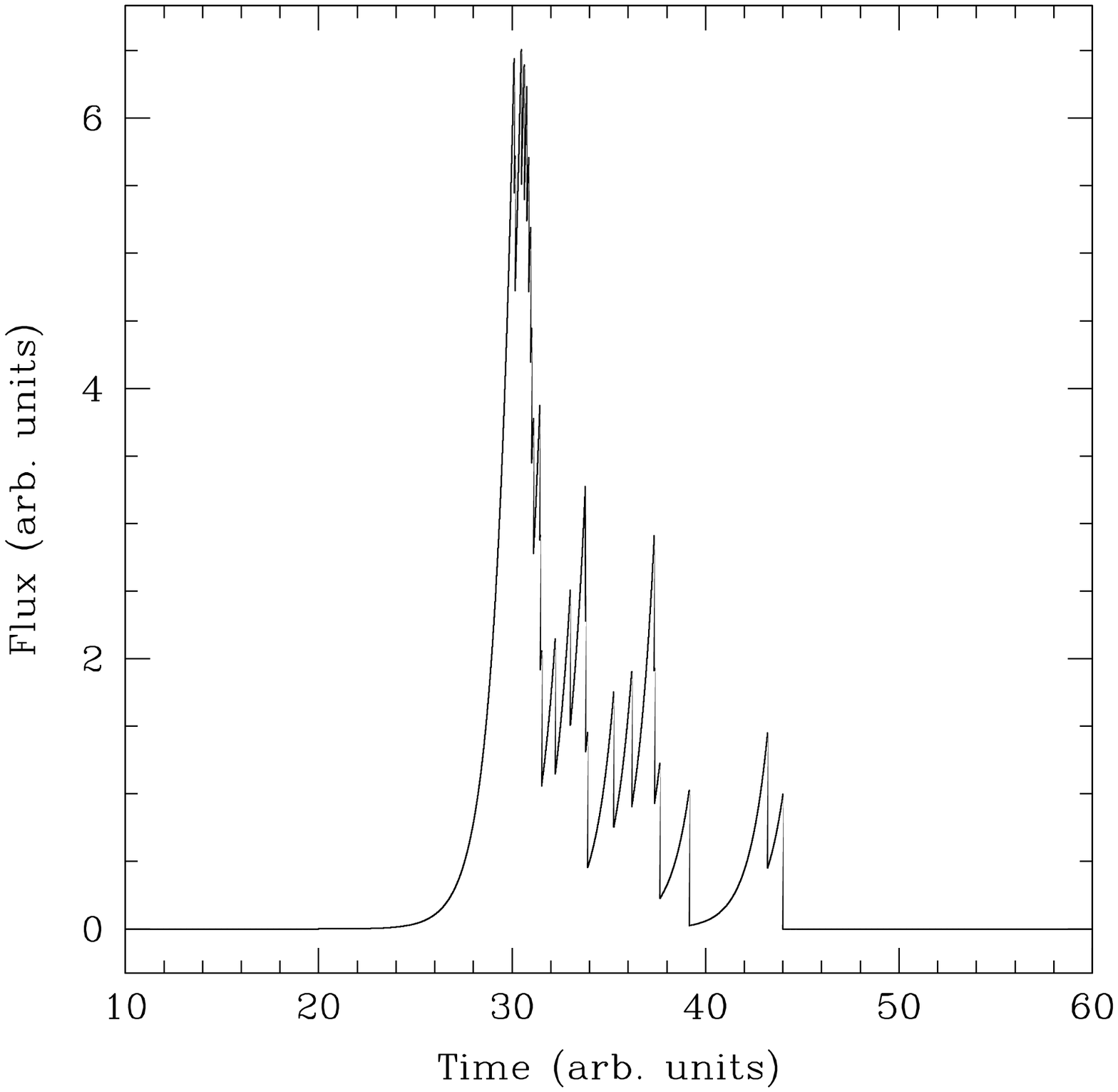}}
\caption{(a)A rising shot (solid line) and a falling shot (dashed
line) (b) Rising shots imbedded in a falling envelope.}
\label{fig:sim_env}
\end{figure*}

An ``envelope'' model with one-sided exponential rising shots with a
rise time of 0.5 seconds and a one-sided exponential envelope with a
decay time of 4 seconds matches the qualitative trends of the observed
skewness, but the overall normalisations of the skewness are a factor
of $\sim10$ too large, indicating that this model assumes much more
asymmetry than is present in nature.  We thus modify the model by
giving both the envelope and the shots two-sided profiles.  The
resulting model lightcurve has a skewness that matches the
observations within factors of a few.  For the time scales on which
the individual shots occur, the skewness is positive.  For time lags
of duration between the individual shot time scale and the
``envelope'' time scale, the skewness is negative.  For time lags
longer than the envelope scale, the skewness is zero.  Skewness
computations for a simulation with shots having an 0.5 second rise
time scale and an 0.1 second decay time scale embedded in an envelope
with a 2.7 second rise time scale, a 3.0 second decay time scale and
53\% of the shots in the decaying portion (so that the rising and
decaying portions of the envelope will form a continuous function) are
computed.  Each model lightcurve contains 25 shots.  Since the process
is random, 149 lightcurves have been computed and the results have
been averaged to improve the statistics.  The simulated skewness for
this model is plotted in Figure 5 (a) and a sample simulated
lightcurve is plotted in Figure 5 (b).  We have also computed the
skewness of a simulated lightcurve made of a random superposition of
the 25 shot lightcurves.  We find that the smearing of the shots
induced by this superposition does not affect the results
substantially.  In order to fit the skewnesses more exactly, one needs
to modify the parameters of the simulation by varying the ratios of
rise time to decay time.  A model in which the shot and envelope time
scales are allowed to vary with some small range around an average
produces essentially the same skewness results as the single time
scale models, and if the number of shots per unit time is too high,
then the individual shots blend together so much that the skewness
reflects only the envelope.  The idea that the shots cannot be
independent stochastic events was also suggested by Uttley \& McHardy
(2001) to explain the constancy of RMS variability as a function of
count rate in Cyg X-1, SAX J1808.4-3658, and several Seyfert galaxies.

A final possibility is related to the third one.  One can have the
shot arrival times be independent of one another, while their
intensities are correlated with one another.  If large shots are
followed only by smaller shots, then the general form of the skewness
also can be reproduced.

\begin{figure*}
\vskip 3 cm \centerline{\epsfxsize=6.2cm \epsfysize=5.33cm
\epsfbox{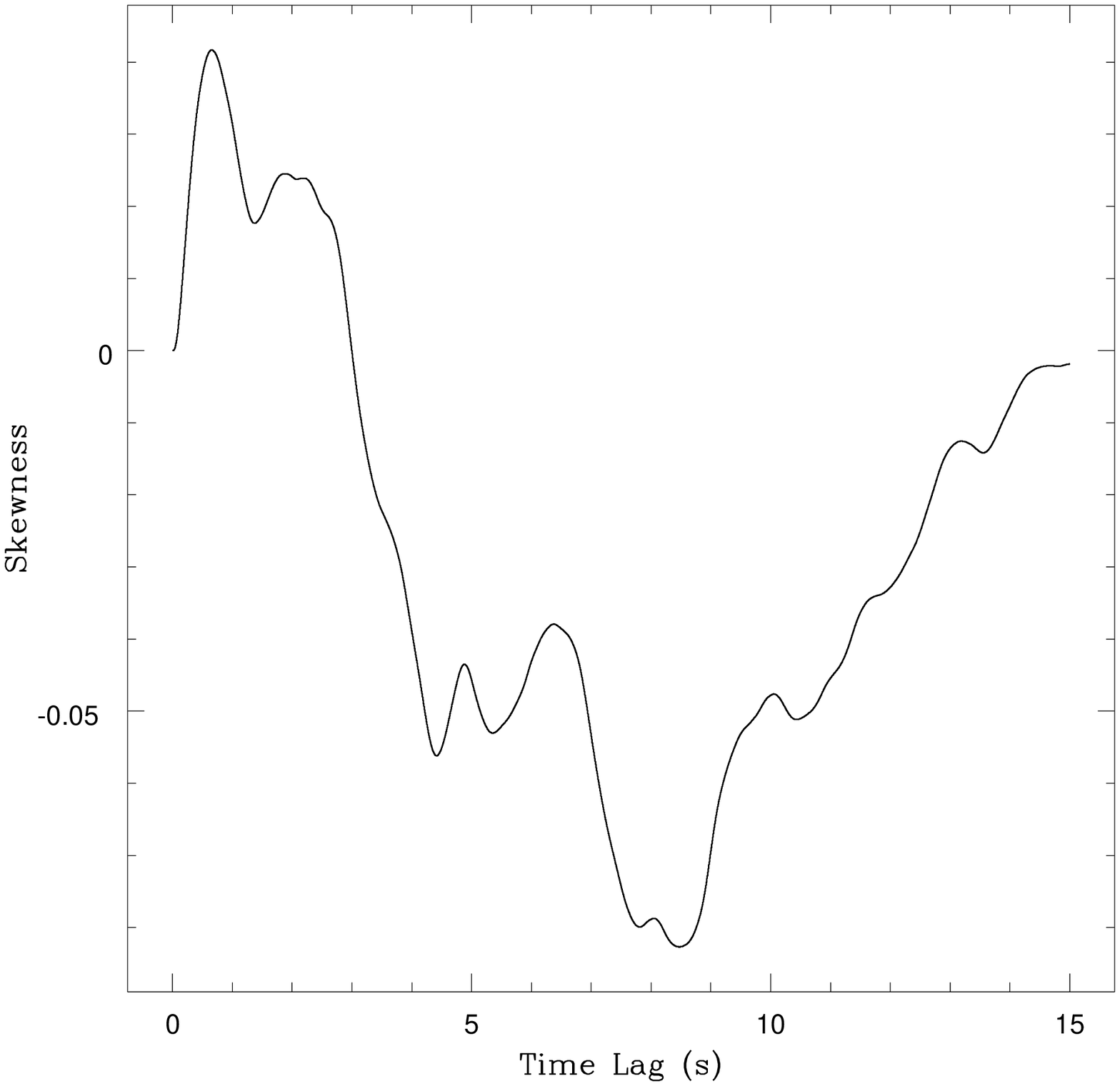} \epsfxsize=6.2cm \epsfysize=5.33cm
\epsfbox{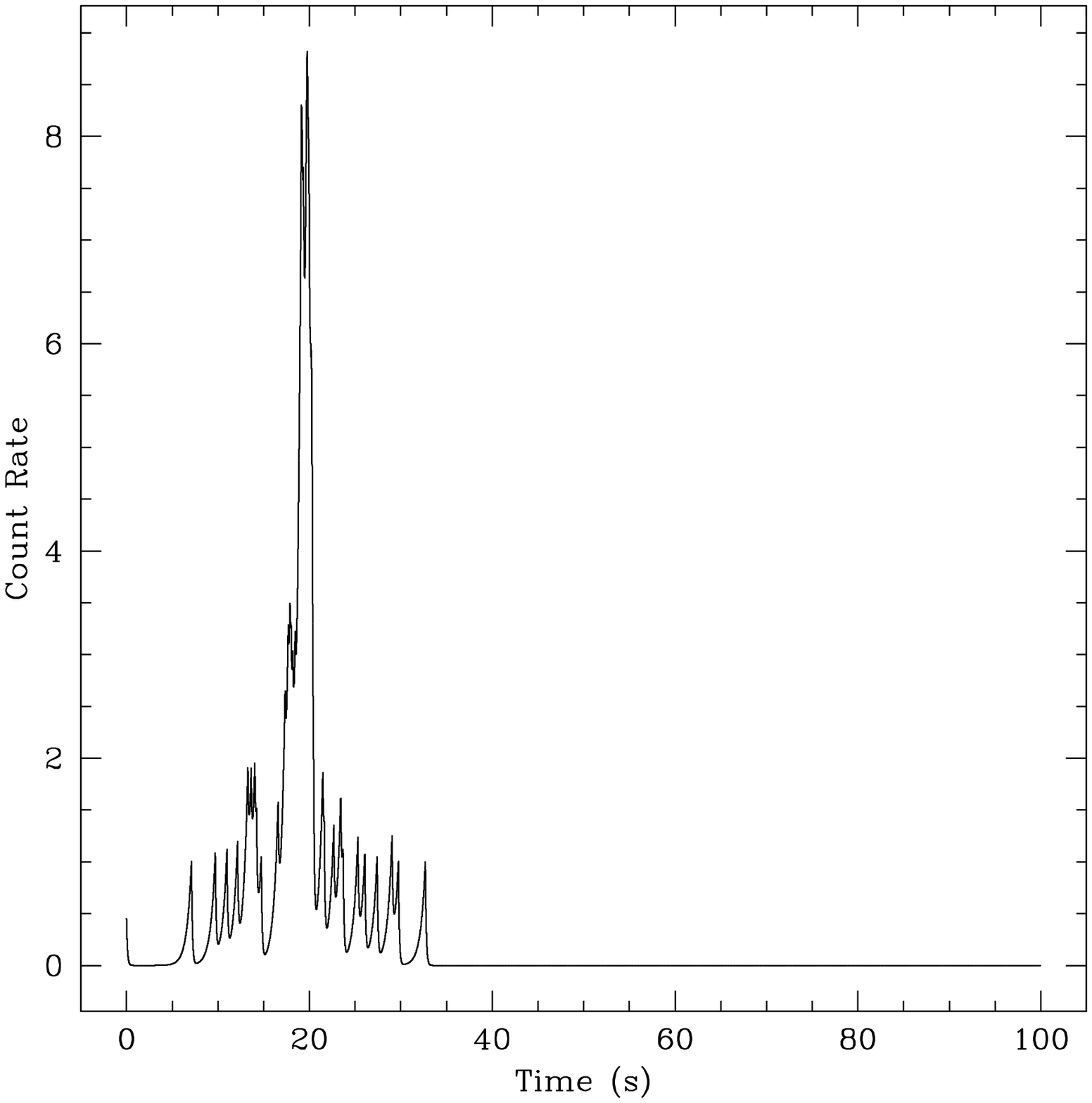}}
\caption{(a)The time skewness of a simulated lightcurve made of an
envelope of shots where the envelope has an exponentially rising
envelope with a 2.7 second time scale, an exponentially decaying
envelope with a 3.0 second time scale and shots with 0.5 second rise
time scales and 0.1 second decay time scales (b) a typical example of
the lightcurve that produced this profile.}
\label{fig:sim_env} 
\end{figure*}

\begin{figure*}
\vskip 3 cm 
\centerline{\epsfxsize=6.2cm \epsfysize=5.33cm \epsfbox{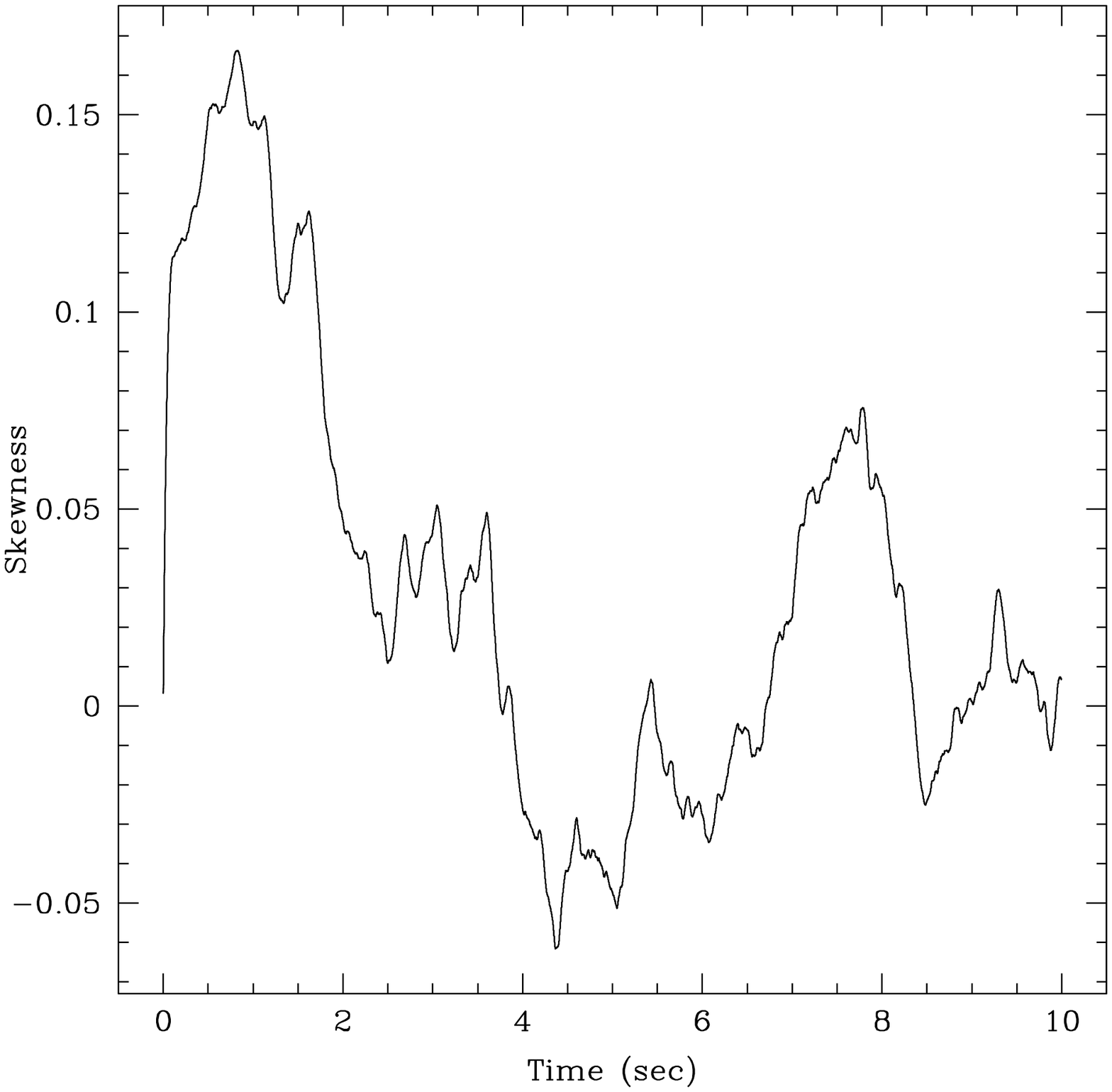}\epsfxsize=6.2cm \epsfysize=5.33cm\epsfbox{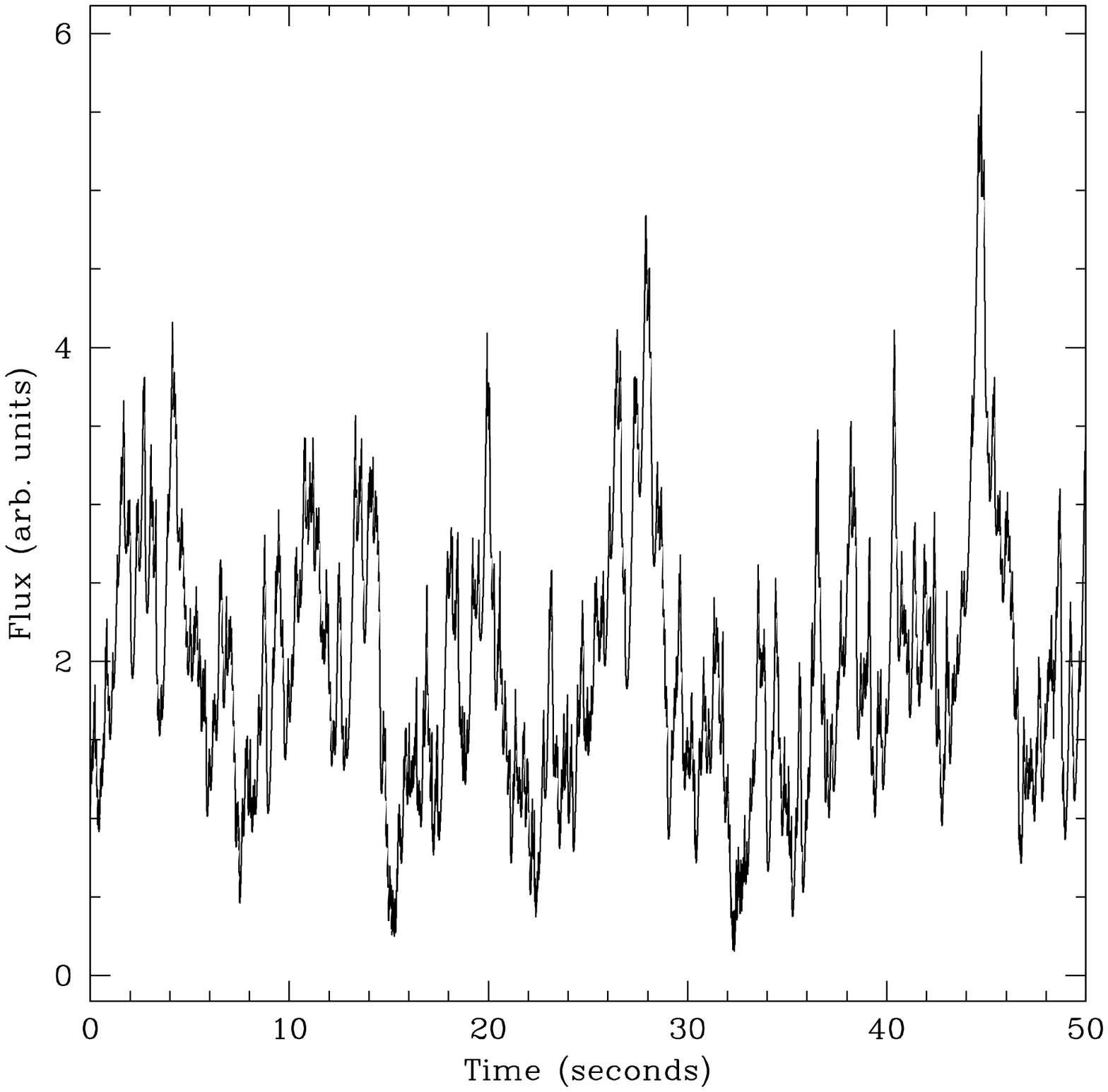}}
\caption{(a)The time skewness of a simulated lightcurve made from a
model in which the size of a given shot is a random number times a
fraction of the energy remaining in an energy reservoir. (b) The first
50 seconds of the simulated lightcurve in our reservoir model.}
\label{fig:sim_soc}
\end{figure*}

Such a mathematical form can be reproduced by two recent models - the
self-organised criticality (SOC) model (Takeuchi, Mineshige \& Negoro,
1995; Takeuchi \& Mineshige 1997) and the pulse avalanche model
(Poutanen \& Fabian 1999).  The skewness of a simulated light curve
that borrows the major results of the SOC model is presented in Figure
6a, and a segment of the simulated light curve is plotted in Figure
6b.  In this simulation, an energy reservoir is posited and shot
arrival times are presumed to be random.  Each shot results in the
release of a random fraction of the energy reservoir.  The duration of
the shot is related to the energy released at $\tau \propto E^{0.4}$.
The energy reservoir is refilled at a constant rate.  Thus when a
large shot occurs, the reservoir's energy level will be depleted and
subsequent shots will be smaller until the reservoir is re-filled.
The shots are given Lorentzian profiles and their $\tau$ values are
given by the ratio of the reservoir's energy level to the mean
reservoir energy level times a random value drawn from a uniform
distribution between 0.05 and 0.15 seconds.  A 1000 second simulated
light curve is produced, allowing us to average over many
reservoir-filling timescales.

Current applications of the SOC model to black hole binary light
curves fail to match several well established observational
constraints for Cygnus X-1 and GX 339-4.  They assume a shot with the
reverse time sense of those seen in the observations.  This is not an
essential component of the model since the shot profile is put in by
hand to match the power spectrum, and our simulation has already
changed this property of the shots and assumed the shots decay rapidly
rather than rise rapidly.  Additionally, this model does not attempt
to explain the time lags seen in black hole binaries in the hard
state.  Both these problems could probably be solved, at least
qualitatively, by assuming that the gas in a shot event gets hotter as
it falls in, then abruptly stop emitting light when it reaches the
event horizon of the black hole.  This would result in a shot with a
rising profile, a sharp turnoff, and hard lags.

More troubling is the inability of the ADAF/SOC model to match the
observed luminosity.  Because it assumes the standard parameters from
Narayan \& Yi (1995), the total amount of energy in electrons in the
corona is small (of order $10^{34}$ to $10^{35}$ ergs), the accretion
rate must be small, and the accretion flow must radiate inefficiently.
This makes it difficult to supply the required luminosity for Cygnus
X-1 and GX 339-4 in their hard states, since the hard states of these
sources are more luminous than those of most other accreting black
hole candidates.  Thus the models of Takeuchi \& Mineshige (1998)
predict a luminosity of only $\sim 10^{34}$ ergs/sec, while the
luminosity of Cygnus X-1 in its hard state is about a factor of 1000
higher than that value.

Still, the skewness of the lightcurve is fairly well matched by the
SOC model.  This success may suggest that some conceptually similar
process - an energy reservoir of some sort drives the variability;
when that reservoir is depleted, production of large shots is
suppressed.  We also note that a gradual mass flow in addition to the
self-organized critical flow is required to produce an $f^{-1}$ power
spectrum rather than a steeper $f^{-2}$ decay.  Whether some form of
the SOC model could reproduce the lightcurve of Cygnus X-1 given a set
of coronal parameters which reproduces the spectrum more accurately
(such as that of Esin et al. 1998) remains to be seen.  {\it At the
present, the problem is with the specific parameters used in the ADAF
realisation of the SOC model and not with the SOC model itself.}  We
note that our simulated ``envelope'' model has a similar problem of
having its power spectrum fall off too rapidly at high frequencies.
To fix the problem, a sharper than exponential envelope function would
be needed.

The magnetic flaring model of Poutanen \& Fabian (1999) is an
alternative mechanism for producing the both the hard emission and the
variability seen in low/hard state objects.  This model produces the
time lags through temperature changes in an active region.  Each shot
has a probability of causing another shot to occur and the mean delay
time before the emission of the stimulated shot is proportional to the
energy of the stimulating shot.  The requirement that the magnetic
energy reservoir be refilled before another shot occurs could be the
cause of this delay.  The possibility that a magnetic energy reservoir
is required to heat the corona in accreting systems has been discussed
recently (Merloni \& Fabian 2001).

Another recent observational result that can perhaps be explained in
terms of these multiple time scales, either due to reservoir effects
or exponential envelopes of shots, is the decomposition of the Fourier
power spectrum into multiple Lorentzian peaks (e.g. Nowak 2000).  This
approach typically yields between 3 and 7 peaks in the power spectrum,
but the basic idea that the variability has multiple characteristic
time scales is present in both our results and the ``many-QPO''
interpretations of the power spectra.  The strongest peaks in the
Fourier spectrum may correspond to the time scales presented here and
the other peaks in the Fourier spectrum may be unresolvable with
current skewness measurements or may require yet higher order moments
of the light curve to be computed.

Finally, assuming that a shot-noise type model is correct, it is noted
that in Cygnus X-1, these results present evidence for a significant
constant (or quasi-constant) flux component.  Since the shots have now
been fairly clearly demonstrated to be shorter at higher energies, the
rms variability expected due to the shots should rise as a function of
energy.  But in the observations (e.g. Nowak et al. 1999), the rms
decreases as a function of increasing energy.  In order to reconcile
these two observations, one must include an additional component with
a significantly harder spectrum than that emitted by the shots, or we
must assume that the shots have different properties at different
energies (e.g. different relative amplitudes or different arrival
frequencies).  In the context of a shot model, the variable relative
amplitude hypothesis would specifically require that the brightest
shots would have softer spectra than the faintest ones, while the
variable arrival frequency would require that some shots show up in
the hard energy bands without contributing to the low energy bands.
Given the observation that the coherence function is greater than 0.95
on all timescales from 0.01 Hz to 10 Hz, but drops significantly above
10 Hz (Nowak et al. 1999), the hypothesis that there are more shots at
high energies than at low energies can be restricted to the case where
the extra shots at high energies contribute almost no power on
timescales longer than 0.1 seconds.  While the near unity
cross-correlation functions can be used to constrain how much the
shots' spectra can vary (Maccarone, Coppi \& Poutanen 2000), rather
large differences in shot spectra can be allowed as long as the peaks
are aligned.  Positing a constant component with a harder spectrum
than the variable component would seem to contradict the results of
Uttley \& McHardy (2001) who found evidence for a constant component
with nearly the same spectrum as the variable component.  Thus it
seems that the picture where the brightest shots have the softest
spectrum is the most likely one.  The possibility that the reflection
component contributes to the drop in rms amplitude at the highest
energies (those $>$ 13 keV, which were not studied by Uttley \&
McHardy) remains viable.

Essentially the same implications are found from the bicoherence
measurements as from the time skewness measurements.  We find that the
$\sim f^{-1}$ dependence of the bicoherence function at high
frequencies cannot be well matched by a single shot model.  In Figure
7 we show the results from a model bicoherence computation for a
simulated lightcurve with an envelope that rises instantaneously and
decays exponentially on a timescale of four seconds, which is composed
of shots that rise on an exponential timescale and decay
instantaneously.  The rise times of the shots are distributed
uniformly between 0.5 and 1.5 seconds.  The resulting bicoherence
measurements are plotted in Figure 7, and match the observations
reasonably well.  Because of the computational intensity of
bicoherence calculations and the uncertainties about whether the shot
shapes should be exponentials, stretched exponentials, or some other
more complicated profile, we cannot find an exact set of parameters
for the shots and their envelopes.

\begin{figure*}
\vskip 2 cm \centerline{\epsfxsize=9.3cm \epsfysize=8.0cm
\epsfbox{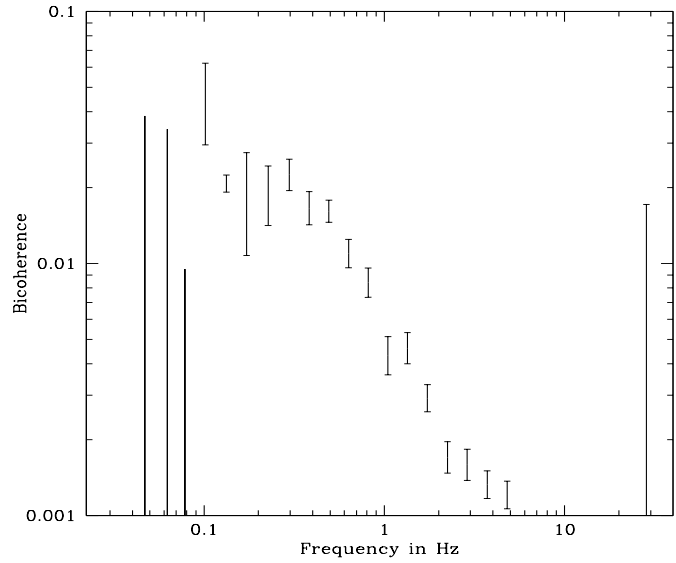}}
\caption{The bicoherence of a simulated light curve with a falling
envelope of rising shots.  The envelope's rise and decay times are 0
and 4 seconds respectively, while the shots' rise times are
distributed randomly between 0.5 and 1.5 seconds and their decay times
are instantaneous.}
\label{fig:sim_soc}
\end{figure*}

\section{Summary} 

We have shown that the hard states of Cygnus X-1 and GX 339-4 have
qualitatively similar skewness and bispectrum properties.  They cannot
be matched by any of the simplest single shot models.  We suggest the
possibility that either a model with a falling envelope of rising
flares or a model with an energy reservoir that suppresses large shots
following other large shots can reproduce the data, and that only
models in which the arrival times and/or luminosities of the shots are
correlated with one another can reproduce the skewness function which
has both positive and negative parts.  Whether the reservoir is one of
accreting mass or of magnetic energy cannot be determined with the
current data.  Of the current variability models, the self organised
criticality models or the magnetic flaring pulse avalanche models seem
most likely to be able to fit this data without major modifications.

\section{Acknowledgments}

We thank Rick Rothschild for pointing out that the time skewness could
be used to test whether shots are symmetric and Roger Blandford for
suggesting the use of the bispectrum to augment the skewness results.
We are grateful to Phil Uttley for useful discussions regarding the
constant component in Cygnus X-1.  We thank the anonymous referee for
making several useful suggestions.  This research has made use of data
obtained from the High Energy Astrophysics Science Archive Research
Center (HEASARC), provided by NASA's Goddard Space Flight Center.

\label{lastpage}

\begin{thebibliography}{}
\bibitem{b3} Beloborodov, A.M., 1999a, in ASP Conf. Ser. 161, High
Energy Processes in Accreting Black Holes, ed. J. Poutanen,
R. Svensson (San Francisco: ASP), 295 (astro-ph/9901108)

\bibitem{} Bullock, T.H., Achimowicz, J.Z., Duckrow, R.B., Spencer,
S.S. \& Iragui-Madoz, V.J., 1995, EEG Clin. Neurophysiol., 103, 661

\bibitem{} Cooray, A., 2001, PhRvD, 64, 3516

\bibitem{} Esin, A.A., Narayan, R., Cui, W., Grove, J.E. \& Zhang,
S.N., 1998, ApJ, 505, 854

\bibitem{b55} Fackrell, J., 1996, Ph.D. Thesis, University of
Edinburgh

\bibitem{b6} Fenimore, E., Madras, C. \& Nayakshin, S., 1996, ApJ,
473, 998

\bibitem{b21} Feng, Y.X., Li, T.P. \& Chen, L., 1999, ApJ, 514, 373

\bibitem{b9} Gierlinski, M., Zdziarski, A.A., Done, C., Johnson, W.N.,
Ebisawa, K., Ueda, Y., Haardt, F. \& Phlips, B.F., 1997, MNRAS, 288,
958

\bibitem{b56} Lin, D., Smith, I.A., B\"ottcher, M. \& Liang, E.P.,
2000, ApJ, 531, 963
 
\bibitem{b1} Maccarone, T.J., Coppi, P.S. \& Poutanen, J., 2000,
ApJL, 537L, 107

\bibitem{b19} Merloni, A. \& Fabian, A.C, 2001, MNRAS, 321, 549

\bibitem{b30} Mendel, J., 1991, Proceedings of the IEEE, 79, 278

\bibitem{b12} Miyamoto, S. \& Kitamoto, S., 1989, Nature, 342, 773

\bibitem{b13} Narayan, R. \& Yi, I., 1995, ApJ, 472, 710

\bibitem{b20} Nowak, M.A., 2000, MNRAS, 318, 361

\bibitem{b21} Nowak, M.A., Vaughan, B.A., Wilms, J., Dove, J.B. \&
Begelman, M.C., 1999, ApJ, 510, 874

\bibitem{b11} Negoro, H., Miyamoto, S. \& Kitamoto, S., 1994, ApJL,
423, L127

\bibitem{b14} Payne, D.G., 1980, ApJ, 237, 951

\bibitem{b18} Poutanen, J. \& Fabian, A. C. 1999, MNRAS, 306, L31

\bibitem{b10} Poutanen, J., 2000, in X-Ray Astronomy 1999: Stellar
Endpoints, AGN, and the Diffuse X-Ray Background, in press
(astro-ph/0002505)

\bibitem{b5} Priedhorsky, W., Garmire, G.P., Rothschild, R., Boldt,
E., Serlemitsos, P. \& Holt, S., 1979, ApJ, 233, 350

\bibitem{b2} Takeuchi, M. \& Mineshige, S., 1997, ApJ, 486, 160

\bibitem{b4} Takeuchi, M., Mineshige, S. \& Negoro, H., 1995, PASJ,
47, 617

\bibitem{b57} Timmer, J. \& K\"onig, M., 1995, A\&A, 300, 707

\bibitem{b40} Uttley, P. \& McHardy, I.M., 2001, MNRAS, 323, L26

\bibitem{b41}
Vaughan, B.A. \& Nowak, M.A., 1997, ApJL, 474, 43L

\bibitem{b7} Zdziarski, A.A., Johnson, W.N., Poutanen, J., Magdziarz,
P. \& Gierlinski, M., 1997, in ESA SP-382, The Transparent Universe,
Proc. of 2nd INTEGRAL workshop, ed. C. Winkler, T. J.-L. Courvoisier,
Ph. Durochoux (Noordwijk: ESA), 373

\bibitem{b8} Zdziarski, A.A., 2000, in IAU Symp.  195, Highly
Energetic Physical Processes and Mechanism for Emission from
Astrophysical Plasmas, ed. P. Martens, S. Tsuruta, in press
(astro-ph/0001078) 

\end{thebibliography}
\end{document}